\documentstyle[prd,aps]{revtex}

\newcommand{\bea}{\begin{eqnarray}}
\newcommand{\eea}{\end{eqnarray}}
\newcommand{\be}{\begin{equation}}
\newcommand{\ee}{\end{equation}}

\begin{document}
\draft
\twocolumn[\hsize\textwidth\columnwidth\hsize\csname
@twocolumnfalse\endcsname

\begin{flushright}
KEK TH-807
\end{flushright}

\title{From Big Crunch to Big Bang: A Quantum String Cosmology Perspective }
\author{Jnanadeva Maharana${}^{(a,b)}$ \\
        ${}^{(a)}$Theory Division,  
                   KEK, Tsukuba, Ibaraki 305-0801, Japan  \\
        ${}^{(b)}$ Institute of Physics, Bhubaneswar-751 005, India}
\maketitle

\begin{abstract}

The scenario that the Universe contracts towards a big crunch and
then undergoes a transition to expanding Universe in envisaged in
the quantum string cosmology approach. The Wheeler-De Witt equation
is solved exactly for an exponential dilaton potential. S-duality invariant
cosmological effective action, for type IIB theory,
 is considered to derive classical
solutions and solve WDW equations. 

\end{abstract}

\vskip2pc]

It is natural to expect that cosmology must be  ultimately founded on quantum
gravity. Since string theory/M-theory synthesise quantum mechanics
and general theory of relativity;  therefore,
the evolution of the Universe in early epochs and fundamental issues concerned
with initial singularity may be resolved in the frame work of string  theory.
 In recent years, there has been considerable amount of
interest in  cosmology from the string theory point of view
\cite{rev1,rev2}. The pre-big bang (PBB)
 scenario
\cite{pbb1},
which has drawn a lot of  attention, proposes an alternative mechanism
for inflation when contrasted with the original paradigm of inflation
 \cite{in1,in2} and promises a possible formulation of nonsingular cosmology.
 One of the postulates of PBB cosmology is that the Universe, in the
remote past, $t\rightarrow -\infty$, is described by weak coupling, low
curvature and cold state and it undergoes an accelerated expansion by the
kinetic energy term of the dilaton while proceeding towards the singularity
lying in its future. Subsequently, there is a transition from the accelerating
to the FRW like  branch in the $t>0$ region. However, one encounters no-go
theorems for the branch change while dealing with the
tree level effective action \cite{gex1,gex2}. There are several other attempts to
understand diverse aspects of cosmology in the frame work of string theory
\cite{tdpol,hm,bbm,dh,bfm}.\\
Recently, Khoury et al. \cite{ko1} have put forward a proposal where the
Universe contracts towards a big crunch and then makes a transition to an
expanding big bang Universe. The scenario envisaged in \cite{ko1} holds the
promise to explore new class of cosmological  models. Furthermore, it has been
pointed out that this idea leads to interesting connection with recently
proposed ekpyrotic model \cite{ko2} which has generated considerable
activities \cite{linde2}. The essential ingredients in ref. \cite{ko1}
is to consider an effective action with graviton and a massless scalar field,
dilaton, describing the evolution of the Universe. This model incorporates
some of the ideas of PBB proposal in that the evolution of the Universe
began in the far past. However;
 it also differs from the scenario of
the PBB that the Universe followed the accelerated expanding branch for 
$t<0$ and then it exits to the expanding, decelerating branch for
which the singularity lies in its past. 
\\
The purpose of this note is to present an investigation of the mechanism
for the transition from big crunch to big bang from the quantum mechanical
perspective. We derive the Wheeler-De Witt equation for the case at hand
and impose appropriate boundary conditions which describes the emergence
of the FRW type Universe starting from initial state which corresponds to
big crunch classically. To be specific, we  adopt an exponential potential
and choose a suitable metric to facilitate  solution for the
case under considerations. We note that {\it raison de etre } for the exponential
potentials have been argued 
 by Moore, Peradze and Salina \cite{mps} from M-theoritic analysis.
Furthermore,  exponential  potentials arising from  
M-theory cosmology might explain quintessence as expounded in \cite{pkt}.
\\
The $D=4$ tree level string effective action is
\bea
S=\int d^4x{\sqrt {-g}} \left({\cal R}_g -{1\over 2}(\partial{\tilde  {\phi}})^2
 -V({\tilde{ \phi}})\right)\eea
Here $g_{\mu\nu}$ is the Einstein frame  metric,
$g$ is its
determinant and ${\cal R}_g$ is the Ricci scalar derived from $g_{\mu\nu}$. 
Note that ${\tilde {\phi}}$ is
the dilaton and $V({\tilde {\phi}})$ is the dilaton potential term.
The metric and dilaton are taken to be time dependent in the cosmological
scenario and we shall consider isotropic, homogeneous and
spatially flat FRW metric.  
In ref \cite{ko1}, the following form of metric was adopted: 
$ds ^2=a(\tau)^2(-N(\tau)^2d\tau ^2 +\sum _{i=1}
^{3}dx_i^2) $, $\tau $ is conformal time. We shall choose
 a different form of metric \cite{jj}
\be
\label{metric}
 ds^2= -{{N(t)^2}\over {{a(t)^2}}}+a(t)^2\delta _{ij}dx^idx^j \ee
Here $N(t)$ is the lapse function.
It is easy to show that 
 ${\sqrt {-g}}{\cal R}_g=- 6N(t) ^{-1} a(t)^2{\dot {a(t)}}^2 +{\rm total ~
derivative ~ term}$. It is very useful for our purpose \cite{jj} to rescale
$a(t)$ and ${\tilde {\phi}}$ to $\phi$ to bring the action to more convenient
form
\bea  L=N(t)^{-1}\left(-{1\over 2}a^2{\dot a}^2+{1\over 2}a^4{\dot {\phi}}^2 
\right)
-Na^2V(\phi) \eea
where the dot denotes time derivative.  The
equation of motion for $N$ leads to  the Hamiltonian
constraint and the equation of motion for $a$  and $\phi$ can be derived easily.
We choose an exponential potential i.e. $V(\phi)=V_0e^{\alpha \phi}$,
$V_0$ being a constant. 
The motivation for choosing the exponential dilaton potential has been alluded
to earlier.  We look for solutions where scale factor,  $a$, has a power law
growth in time: $a(t) =|t|^p, ~~p>0$.
The Hubble parameter, H, its time derivative, the dilaton, $\phi $ and its
time derivative have the following form as can be inferred from the equations
of motion. 
\be H={{\dot a}\over a}={p\over t},~~~{\dot H}=-{p\over {t^2}} \ee
\be
\phi ={\phi}_0 +{\sqrt{{{(p(1-p)}\over 3}}} ln|t| , ~~{\dot {\phi}}={\sqrt{{{
p(1-p)}\over 3}}} {1\over t} \ee
 Here $a_0$ and ${\phi}_0$ are arbitrary constants. The parameter $\alpha$
appearing in the potential gets fixed in terms of exponent $p$ and so is the
ratio of $V_0$ and $a_0 ^2$,
\be \alpha =-2{\sqrt{{{3(1-p)}\over p}}}, ~~~{\rm and}~~~~~{{V_0}\over {a_0 ^2}}
={{4p^2-p}\over 6} \ee
Cosmological solutions with exponential potentials, for different form  
of metric, has been considered in the  past \cite{past} and the have been
topics of discussion more recently in the context of ekpyrotic model
\cite{present}. Note that, $G_N$ or the Planck mass, $m_P$ does not appear
in our action and therefore, these constants are also absent in our solutions
of equations of motion. From now on we {\it choose} $\alpha =-2$ so that
\be V=V_0e^{-2\phi}, ~~~~{\rm correspondingly}~~~~~~~~p={3\over 4} \ee
Note that  we have also absorbed a factor of $e^{-2\phi _0}$ in the
definition of $V_0$. We have mentioned earlier the justification
for adopting an exponential form of potential; however, it is possible that
 this
form of potential might not be adequate all the way
 to very small values of $t$. 
Let us define, new
set of variables: $ u={{a^2}\over 2}{\rm cosh}2\phi$  and
$ v={{a^2}\over 2}
{\rm sinh}2\phi$; then the Lagrangian is
\be L={1\over {2N}}\left({\dot v}^2-{\dot u}^2\right) -{N\over 2} V_0(u-v) \ee
for our choice of specific exponential potential.
The corresponding Hamiltonian is
\be
\label{hamil}
{\cal H}= N ({1\over 2} p_v^2 -{1\over 2}p_u^2
 +{{V_0}\over 2}[u-v] ) \ee
varying $N$ yields the constraint $H=0$ in $N=1$ gauge. The momenta
$p_v$ and $p_u$ are conjugate to  $v$ and $u$ respectively. Furthermore, 
$[p_u+p_v, {\cal H} ]_{PB}=0$, implying existence of a conserved charge,
\be 
\label{qch}
Q={{e^{-2\phi}\over {a(t) ^2}}}[-H+{\dot \phi}^2] \ee
 $H$ is the Hubble parameter.  
Note that $\phi =\phi _0 +{1\over 4}ln |t|$, where $\phi _0$ is
a constant which is value that dilaton assumes at some constant, $t=t_o$.
Thus  Q contains a 
 factor 
 $e^{-2\phi _0} $ in its definition, revealing the  coupling constant
 dependence. 
\\ 
Now we obtain the Wheeler-De Witt equation and 
impose appropriate boundary conditions on the  wave function. It has been
advocated that quantum string cosmology might be useful to address the issue
of graceful exit \cite{wgmv,mmp,mu}and to study evolution of the early 
Universe  
\cite{jim12}. 

The
Wheeler-De Witt equation takes the following form  for the Hamiltonian (set $N=1$)
(\ref{hamil}), 
\bea 
\label{weq}
[ -{{\partial ^2}\over {{\partial} v ^2}}+{{\partial ^2}\over {{
\partial }u^2}}+{{V_0}\over 2}(u-v)]\psi (v,u)=0 \eea
We may solve (\ref{weq}) by  separation of  $v$ and $u$
and $\psi (u,v)$ is  product of two  Airy functions.
However, it is useful to reexpress the WDW equation 
in terms of another set of variables and
the wave function thus derived has a more direct intepretation in terms
of a scenario that the Universe evolved towards (classical path of) big
crunch in negative $t$ region and then undergoes a transition to the
expanding phase for $t>0$. Define, 
\be \xi={1\over 6}ln4XY ,~~~ \zeta={1\over 6}ln {X\over Y} \ee
with  $X={1\over 4}(u+v)$ and $Y={1\over 8}(u-v)^2$. The Wheeler-De Witt
equation is expressed as  (in the gauge $N^{-1}= {1\over 2}[u-v]$ )
\be
{\cal H} \psi (\xi , \zeta)=\left({{\partial ^2}\over {{\partial}\xi ^2}} 
- {{\partial
^2}\over {{\partial }\zeta ^2}} +9V_0e^{{6\xi}}\right) \psi (\xi ,\zeta) =0 \ee
The wave function is a product
of a plane wave in $\zeta$ variable of the form $e^{{\pm}ik\zeta}$, where
 $k$ is the separation constant. Note that $k$ is
also identified with eigenvalue of the momentum operator $i {{\partial}\over
{{\partial }\zeta}}$ acting on the plane wave solution (see the arguments
in \cite{wgmv} for the choice of the sign in defining operator $p_{\zeta}$)
 and this is a
constant of motion since $[p_{\zeta} ,H]=0$. Indeed, the  conserved momentum is 
  related to the charge $Q$, defined in (10). 
The solution to the equation of  $\xi$ variable
is Bessel function; therefore,  $\psi$ is given by
\be \psi _k(\xi ,\zeta)= e^{{\pm } ik\zeta}{\cal F} _{\pm {{ik}\over 3}}(z) ,~~~~
z={{\sqrt {V_0}}}e^{3\xi}  \ee
where ${\cal F} _{\nu} (z)$ is one of the Bessel 
functions \cite{abst}, $J_{\nu}(z)$, $Y_{\nu}
(z)$ or Hankel functions $H_{\nu}^{(1,2)}(z)$. The relevant Bessel
function is chosen in order to fulfill desired boundary conditions. 
We demand that there is only right moving wave in the negative $t$ region, with
positive eigenvalue of $p_{\xi}$,
and are led to choose $J_{-i{k\over 3}}(z)$. In this domain, for small geometries,
as $a \rightarrow 0$, 
$z \rightarrow 0$  and
\be {\rm lim}_{z \rightarrow 0} ~~~J_{-i{k\over 3}}(z) \sim {1\over 2}(z)^{-i
{k\over 3}} \sim e^{-ik\xi} \ee
Thus, in this limit, the wave function behaves as 
\be
\psi _k(\xi ,\zeta) \sim e^{-ik(\xi +\zeta)}  \ee
 The Universe expands in the $t>0$ region. The scale factor grows
with time and we are led to consider behavior of $\psi$  when 
$\xi \rightarrow \infty$. 
The asymptotic form of $\psi$ is expressed as sum of two components: 
\bea
\psi _k(\xi ,\zeta)=e^{-ik\zeta}J_{\nu}(z)=_{\xi  \rightarrow \infty }
\psi _{k(+)} +\psi _{k(-)}    \eea  
where $\nu =-i{k\over 3}$ and the two components are given by   
\be
\psi _{k(\pm)}= {\sqrt {{2\over {\pi z}}}}e^{-ik\zeta}e^{\mp i(z- {\pi \over 2}
-{1\over 2}\nu \pi -{\pi \over 4})}
\ee
Moreover,  $\psi _{k(\pm)}$ satisfy following relations when
operated up on by the momentum operator,
 $p_{\xi} =-i{\partial \over {\partial \xi}}$
\be p_{\xi}\psi _{k(\pm)}=\mp z\psi _{k(\pm)} \ee
Thus, the choice of the wave function describes the evolution of the Universe
as follows: 
 the right moving wave propagates from the -ve $t$ region.
The wave function in the  positive time domain is superposition of the left and
right moving components. The boundary conditions adopted to choose the
wave function is in the spirit of Vilenkin's proposal \cite{vil}. 
The probability for the transition  
from  the branch with wave function of the
form $e^{-ik\zeta -ik z}$ to the branch which has wave function
$e^{-ik\zeta +ik z}$, in the asymptotic limit, is given by 
\be P_k={{|\psi _{k(-)}|^2}\over {|\psi _{k(+)}|^2}}=e^{-{2\over 3}\pi k} \ee
Recall that $k$ is the conserved momentum. Note that the
the  conserved charge (\ref{qch}) arises due to conservation of momentum.
Therefore,  $k = {\rm const ~ e}^{-2\phi _0}$ and we are not in a position to
determine the exact value of the multiplicative constant; although our   
arguments show how a factor of $e^{-2\phi _0}$ appears in definition of $k$. 
Thus the probability gets
suppressed for in the weak coupling phase. \\
Next, we proceed to discuss cosmological solutions, in the present context,
for graviton-dilaton-axion system which appear in the 
type IIB effective action. 
The
dilaton, $\phi$ and RR scalar, $\chi$, called axion, belong to $SL(2,R)$
S-duality group and they parametrize the coset ${SL(2,R)}\over {SO(2)}$. \\
The relevance of this problem in the present context 
can be perceived as follows. The presence
of axion introduces the kinetic energy term $a^4e^{2\phi}{\dot{ \chi}}^2$, 
in  the Einstein frame,  in
the action besides the curvature scalar and kinetic energy term  
of the dilaton. When, one considers the $SL(2,R)$ invariant for of the
type IIB action, it is not possible to introduce a nontrivial $SL(2,R)$
invariant potential (depending on dilaton and axion) as has been argued  
in \cite{sdu}. Thus, the equation of motion for the axion is a conservation
law i.e. there is a conserved charge $Q_a= a^4e^{2\phi}{\dot {\chi}}$. 
Therefore, while solving coupled set of equations of motion, we may
eliminate the kinetic energy term of axion ( or ${\dot {\chi}}$ wherever it
appears)
and end up with a term which looks like an exponential dilaton potential.
Furthermore, while solving the Wheeler-de Witt equation, we solve the
wave equation with scale  factor, dilaton and a potential depending on
scale factor and exponential of dilaton. However, it is much more
convenient to express the type IIB action in terms of $SL(2,R)$ matrices
so that the action is manifestly S-duality invariant. Therefore,
I shall adopt the form of the action which could be obtained through
toroidal compactification of ten dimensional type IIB action to four
dimensions as was shown by me a few years ago \cite{sdu}.\\ 
The 10-dimensional effective action can expressed in manifestly S-duality
invariant form \cite{s1} in the Einstein frame. The  toroidal
compactification of that action to lower dimensions, preserving S-duality
invariance, was presented by the author and Roy \cite{sdu,roy}   
which has been useful to obtain classical solutions of IIB theory. 
 We consider a simple version of the
 4-dimensional type IIB effective action: the compactification 
radii of the tori are taken to be constant (set to one), only graviton, 
dilaton and
axion are retained in the reduced action and furthermore, rest of the
scalar, vector and tensor fields are set to zero. We refer the interested 
reader 
to \cite{sdu} where the full reduced action is derived. Our starting point
is 
\be S_4=\int d^4x{\sqrt {-g}}\left({\cal R}_g +{1\over 4}{\rm Tr}[\partial
_{\mu}{\cal M}\Sigma {\partial ^{\mu}}{\cal M} \Sigma ]\right) \ee
This  is the action in the Einstein frame  with, 
\be {\cal M}=\pmatrix {\chi ^2e^{\phi}+e^{-\phi} & \chi e^{\phi} \cr
\chi e^{\phi} & e^{\phi}\cr } , ~~{\rm and}~~\Sigma=\pmatrix {0 & i \cr -i & 0
\cr} \ee
Here $\Sigma$ is the SL(2,R) metric in our conventions \cite{sdu}. 
The above action is manifestly invariant under S-duality transformations
\be {\cal M}\rightarrow \Lambda {\cal M}\Lambda ^T,~~g_{\mu\nu}\rightarrow
g_{\mu\nu},~~~\Lambda \Sigma \Lambda ^T=\Sigma \ee
where $\Lambda \in SL(2,R)$  with unit determinant.
Some of the useful relations these matrices satisfy  are
\be \Sigma \Lambda \Sigma =\Lambda ^{-1}, {\cal M}\Sigma {\cal M}=\Sigma ,~~and
~~\Sigma {\cal M}\Sigma ={\cal M}^{-1} \ee
Note that ${\cal M} \in SL(2,R)$ and is symmetric. For the cosmological case,
with our form of FRW metric (after some rescaling) the
Langrangian is
\be 
\label{lm}
L=-{1\over {2N}}a^2{\dot a}^2 -{{a^4}\over 4N}{\rm Tr}[{\dot {\cal M}}\Sigma
{\dot {\cal M}}\Sigma ] \ee
$\cal M$-equation of  motion deserves some care, since ${\cal M}\in
SL(2,R)$ and is  a conservation law as expected \cite{sdu,masv}.
\be 
\label{meq}
\partial _0({\sqrt {-g}}g^{00}{\cal M}\Sigma \partial _0{\cal M}\Sigma)=0
\ee
and thus we conclude
$ {\sqrt {-g}}g^{00}{\cal M}\Sigma \partial _0{\cal M}=A $, 
where $A$ is a constant $2\times 2$ matrix; is  symmetric  and satisfies
$A\Sigma {\cal M}=- {\cal M}\Sigma A$ which follow by using the relations
between ${\cal M}$ and $\Sigma$ mentioned above. Another important relation is
(using $g^{00}{\sqrt {-g}} =a^4$)
\be 
\label{trace}
{\rm Tr}({\dot{\cal M}}\Sigma {\dot {\cal M}}\Sigma) = -{1\over {a^8}}
{\rm Tr}(A\Sigma A\Sigma) \ee
 The time dependence in the $r.h.s.$
is buried in the presence of $a$. The Hamiltonian constraint 
relates the Hubble parameter and scale
factor appearing through relation (\ref{trace})  
\be
H^2={1\over {2a^8}}{\rm Tr}(A\Sigma A\Sigma) \ee
resulting in the  time dependence: $a=a_0t^{1\over 4} $, $a_0$
being a constant including the factor coming from Tr$A\Sigma A\Sigma$. Notice that 
 the  Einstein-Friedman equation derived from (\ref{lm}) 
is also satisfied when (28) solved.  We can solve (\ref{meq})
once  $a(t)$ is determined  
\be {\cal M}(t)=e^{A\Sigma ln(t-t_0)}{\cal M}(t_0) \ee
where $t_0$ is an arbitrary constant and ${\cal M}(t_0)$ is value of the matrix
at $t_0$. 
\\
Let us derive the WDW equation for the $SL(2,R)$ invariant system. First of all
define $a(t)=e^{\alpha (t)}$ and then a new time variable through
the relation $d\tau =e^{-4\alpha}dt$ and the derivatives with respect to $\tau$
are denoted by prime.
\be L=\int d\tau [-{1\over 2}{\alpha '} ^2 -{1\over 4}
{\rm Tr}({\cal M}'\Sigma {\cal M '}\Sigma)] 
\ee 
$\alpha '$ appearing here not to be confused with inverse string tension. The
canonical momenta are
\be p_{\alpha }=-\alpha ' ,~~~{\Pi}_{\cal M}= - {1\over 2}\Sigma{\cal M}'
\Sigma \ee
The canonical Hamiltonian is
\be {\cal H}=
=-{1\over 2}p_{\alpha}^2 -{\rm Tr}(\Sigma \Pi _{\cal M}\Sigma \Pi _{\cal M}) \ee
The 
WDW equation assumes the following  form
\be
\left({{\delta ^2}\over{\delta{\alpha ^2}}}+{\rm Tr}(\Sigma{{\delta}\over{
\delta {\cal M}}}\Sigma {{\delta }\over{\delta {\cal M}}}) \right)\Psi(\alpha ,
{\cal M})=0 \ee
We may factorize $\Psi (\alpha ,{\cal M})={\cal F}(\alpha){\cal G}(\cal M) $.    
From the conservation law of the $\cal M$-matrix evolution equation,  
the quantum mechanical relation
\be {\cal M}\Pi _{\cal M} {\cal G}=i{\cal M}{{\delta}\over {\delta {\cal M}}}
{\cal G}=(2A\Sigma){\cal G} \ee
follows immediately. The wave equation satisfied by $\cal F$ is
\be \left({{\delta ^2}\over{ \delta {\alpha ^2}}}+{1\over 4}{\rm Tr}(A\Sigma)^2 
\right){\cal F}(\alpha)=0 \ee
which leads to a 'plane wave' solution
\be {\cal F}(\alpha)=e^{\pm i{{\alpha }\over 2}[{\rm Tr}(A\Sigma)^2]^{1\over 2}}
\ee
To solve for $\cal G$ with constraint (35), one needs to specify the matrix
 $A$. For example, when $\chi =0$, $\cal M$ is diagonal and $\cal G$
is pure dilatonic plane wave. More general form of A will lead to 
interesting class of solutions respecting S-duality. 
An important point is  that an arbitrary 
potential $V(\phi)$ added to Langrangian (\ref{lm}) breaks S-duality
symmetry; moreover, the choice of S-duality invariant potentials are severely
restricted \cite{sdu}. Therefore, such symmetry considerations might play 
important roles in the study of cosmological solutions \cite{edc}.\\
It is worth while to recall the scenario proposed in \cite{ko1} where
the Universe evolves from large negative $t$ towards the big crunch.
 Subsequently it undergoes a transition to a FRW-like branch in the
$t\ge 0$ domain. When we contrast this 
proposal with the PBB \cite{pbb1} picture,
the Universe proceeds  towards the singularity as it accelerates in the
$t<0$ domain. Then, it is expected to go through a transition to expanding,
decelerating phase for positive $t$. These two branches are related
to each other through scale factor duality symmetry. In the  proposal,
where the Universe is proceeding towards big crunch, it also approaches
weak coupling regime and therefore, the perturbation theory could be trusted in
that neighborhood.
Of course, there might be additional terms which could give additional  
contributions near the singularity. In this investigation of quantum
string cosmology, we do not take into account those effects. 
\\
In summary, we have presented a quantum string cosmological investigation
of the scenario that the intial state of the Universe is the one which evolves
towards big crunch in negative regime and subsequently, the wave function
gets reflected. The probability is exponetially suppressed in the weak coupling
regime.At this stage, this suppression is shown to be true for
the typical potential we have chosen here; however, this result might
be valid under general conditions although no such proof exists at
the moment.  Furthermore, we considered an S-duality invariant action and solved
the classical equations in a general setting and discussed the structures of
the wave functions of the corresponding WDW equations.\\
I would like to thank Professor Y. Kitazawa for 
his interests and suggestions. It
is a pleasure to acknowledge gracious hospitality of Professor Kitazawa and
the KEK.

\bigskip


\end{document}